\documentclass[aps,pra,twocolumn,showpacs,superscriptaddress,nofootinbib]{revtex4}

\usepackage{amsmath}
\usepackage{amsfonts}
\usepackage{amssymb}
\usepackage{mathdots}
\usepackage{graphicx}
\usepackage{bbm}
\usepackage{color}
\usepackage{pifont}
\usepackage{enumerate}
\usepackage{booktabs}
\usepackage{multirow}

\DeclareMathOperator{\tr}{tr}

\newcommand{\ket}[1]{|#1\rangle}
\newcommand{\bra}[1]{\langle#1|}

\newcommand{\ketbra}[2]{\ket{#1}\bra{#2}}

\newcommand{\MA}{\mathbf M_A}
\newcommand{\MB}{\mathbf M_B}

\newcommand{\MAu}{\mathbf M_A^u}
\newcommand{\MBu}{\mathbf M_B^u}

\newcommand{\rh}{\rho_{AB}}
\newcommand{\sig}{\sigma_{AB}}

\newcommand{\barrh}{\bar \rho_{AB}}
\newcommand{\even}{\mathrm{even}}
\newcommand{\odd}{\mathrm{odd}}

\newcommand{\Fu}{\mathcal F^u}
\newcommand{\F}{\mathcal F}
\newcommand{\Ug}{\mathcal U_g}

\newtheorem{definition}{Definition}

\begin{document}

\title{Security proof of the unbalanced phase-encoded BB84 protocol}

\author{Agnes Ferenczi}
\email[]{aferenczi@iqc.ca}

\affiliation{Institute for Quantum Computing \& Department for Physics and Astronomy, University of Waterloo, 200 University Avenue West, N2L 3G1, Waterloo, Ontario, Canada }
\author{Varun Narasimhachar}
\affiliation{Institute for Quantum Computing \& Department for Physics and Astronomy, University of Waterloo, 200 University Avenue West, N2L 3G1, Waterloo, Ontario, Canada }
\author{ Norbert L\" utkenhaus}
\affiliation{Institute for Quantum Computing \& Department for Physics and Astronomy, University of Waterloo, 200 University Avenue West, N2L 3G1, Waterloo, Ontario, Canada }

\date{\today}
\pacs{03.67.Dd, 03.67.Hk, 42.50.Ex,  42.79.Sz}


\begin{abstract}

In optical implementations of the phase-encoded BB84 protocol, the bit information is usually encoded in the phase of two consecutive photon pulses generated in a Mach-Zehnder interferometer. In the actual experimental realization, the loss in the arms of the Mach-Zehnder interferometer is not balanced, for example because only one arm contains a lossy phase modulator. Therefore, the amplitudes of the pulses is not balanced, and the structure of the signals and measurements no longer corresponds to the (balanced) ideal BB84 protocol. Hence, the BB84 security analysis no longer applies in this scenario. 
We provide a security proof of the unbalanced phase-encoded BB84. The resulting key rate turns out to be lower than the key rate of the ideal BB84 protocol. Therefore, in order to guarantee security, the loss due to the phase modulator cannot be ignored. 

\end{abstract}

\keywords{Quantum key distribution with practical devices, Phase-encoded BB84}

\maketitle

Quantum key distribution (QKD) provides a way for two distant parties (Alice and Bob) to establish a shared secret key with absolute confidentiality. Many protocols \cite{scarani09a} have been suggested to achieve this goal, among which the BB84 protocol \cite{bennett84a} is the most well-known example. 
In the BB84 protocol, Alice randomly chooses between two conjugate bases of a qubit Hilbert space, and encodes the bit value of the key elements in the basis states. She sends these states to Bob through a quantum channel, who measures them randomly in one of the conjugate bases.  After having collected enough data, 
they perform error correction to eliminate the errors in their data, followed by privacy amplification to guarantee the security of the generated key from an eavesdropper (Eve).

In optical implementations, the bit information is usually encoded in a photonic degree of freedom, e.g., in the polarization of photons, or the phase of two consecutive photon pulses. In the phase-encoded protocol, the phase between two consecutive pulses prepared by Alice determine the bit and the basis value of the sent signal. 
In the actual experimental realization of the phase-encoded BB84 protocol with Mach-Zehnder interferometers (see Fig. \ref{setup_fig}), the phase modulator, which is in one arm of the interferometer, introduces loss. 
While this does not change the observed error rate in the data, it changes the signal states and the measurements of the protocol. 
Since this is now a different protocol, the security proofs tailored to the BB84 protocol no longer apply in this scenario. 

In this work, we provide a security proof of the phase-encoded BB84 where we take into account the loss in the phase modulator (``unbalanced phase-encoded protocol''). We use the security approach presented in Refs. \cite{devetak05a, kraus05a} to calculate the key rate. This security approach is valid when Eve is restricted to collective attacks, but in many situations, it also holds for the more general coherent attacks.

We provide a qubit-based security proof, that is later extend to optical modes. At the source side, we extend the validity of the qubit-based security proof to optical modes using the tagging approach introduced in Refs. \cite{gottesman04a, inamori07a} in the decoy framework \cite{hwang03a, lo05a, wang05a}. On the other hand, the squashing model in Refs. \cite{beaudry08a, tsurumaru08a, narasimhachar11asub} justifies the assumption of a qubit-based security proof at the receiver's side.


\section{Protocol setup and proof techniques}

\subsection{Unbalanced phase-encoded protocol}

\begin{figure*}
\includegraphics[width=1.5\columnwidth]{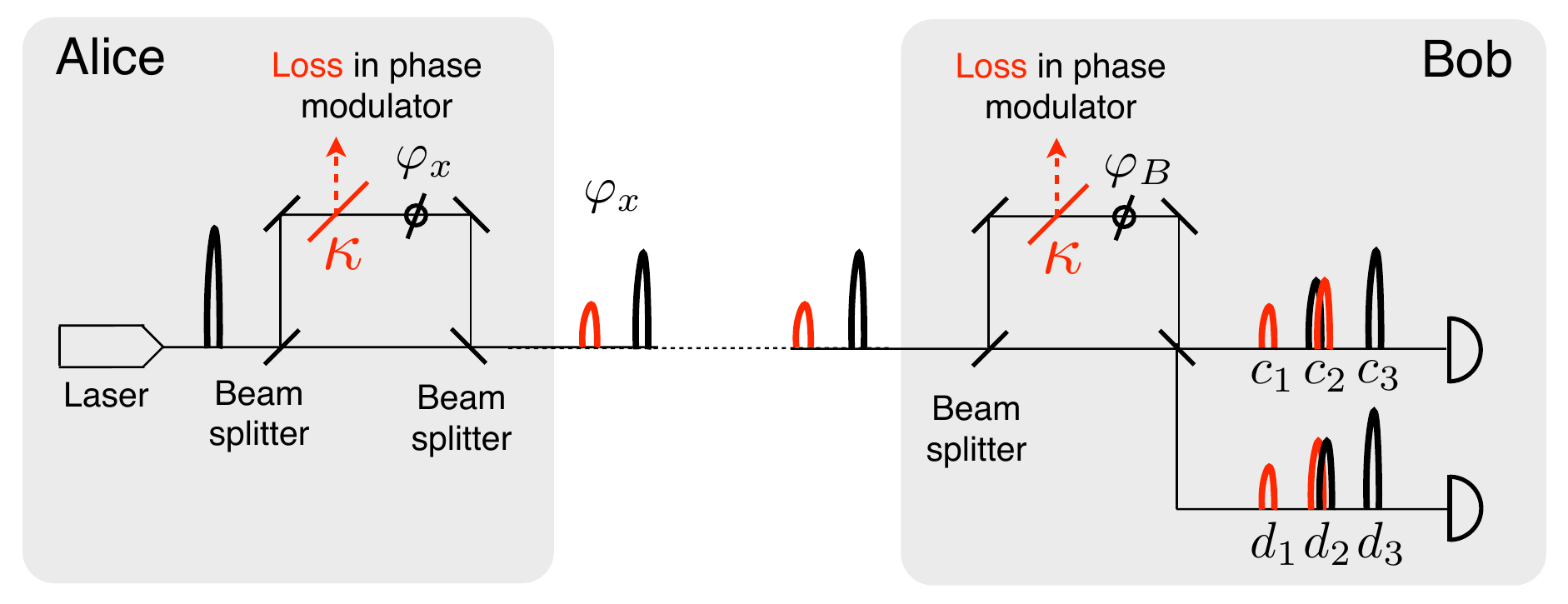}
\caption{ \label{setup_fig} Alice and Bob use a Mach-Zehnder interferometer to prepare and detect the signal pulses. Only the interfering pulses, which produce clicks in the time slots $c_2$ and $d_2$ (black-red and red-black overlapping pulses) are used for the key generation. }
\end{figure*}

A scheme of the protocol setup is provided in Fig. \ref{setup_fig}. 
Alice sends photon pulses through a Mach-Zehnder interferometer with a long arm and a short arm, to create the signal states.
In the long arm, Alice changes the relative phase $\varphi_x$ of the two pulses with a phase modulator to imprint the basis and the bit information on the signal.
Alice chooses the phases $\varphi_x = \frac{\pi}{2} x$ for $x \in \{0, 1, 2, 3 \} $ with equal probability for the 4 signal states. The phases $\varphi_x \in \{ 0 , \pi \}$ and $\varphi_x \in \{ \pi/2, 3\pi/2 \} $ correspond to the bit values $\{ 0, 1\}$ in the ``even'' and ``odd'' basis.
Likewise, the receiver (Bob) detects the signals by means of a Mach-Zehnder interferometer. Bob chooses the phase $\varphi_B \in \{ 0, \pi/2 \}$, which determines the basis (``even'' or ``odd'') of his measurement. Bob chooses each measurement setting with probability $1/2$. 

The pulses arrive in Bob's detectors in three different time slots, either in the top output port (slots $c_1$, $c_2$ and $c_3$ in Fig. \ref{setup_fig}) or in the bottom output port (slots $d_1$, $d_2$ and $d_3$ in Fig. \ref{setup_fig}). Only the middle clicks (slots $c_2$ and $d_2$) are used for the key generation. The outside clicks (slots $c_1$, $c_3$, $d_1$ and $d_3$) are pulses that did not interfere at Bob's second beam splitter. 
If the signal produces interference (e.g. the detectors click in the middle time slot), then Bob determines the bit value of the incoming signal based on his phase setting.

To obtain the data for the key generation, Alice and Bob apply a sifting step, in which they announce to each signal the basis (``even'' or ``odd'') publicly and discard all data points where the basis did not match.  

The lossy phase modulator typically introduces a loss in one of the arms of the interferometer, producing pulses with different amplitudes. We model the lossy phase modulator by a perfect (lossless) phase modulator followed by a beamsplitter with transmissivity $\kappa \leq 1$ that simulates the loss.

\subsection{PBS protocol}

\begin{figure} 
\centering
\includegraphics[width=0.6\columnwidth]{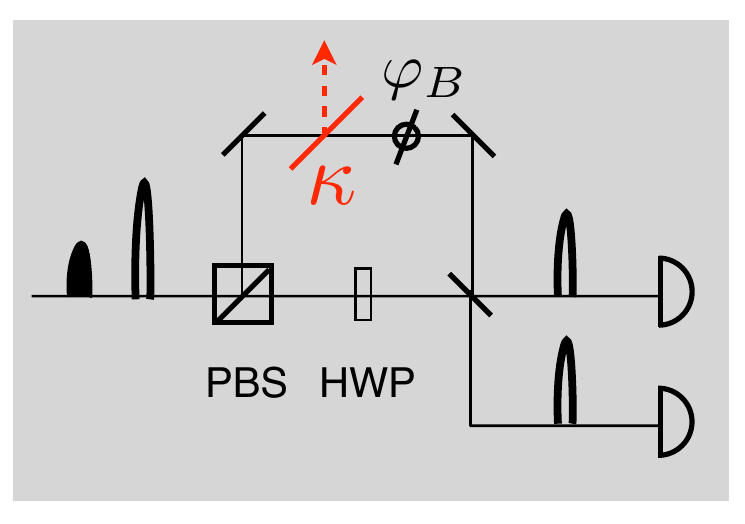}
\caption{ \label{pbs_fig} A variation of the protocol with the pulses encoded in different polarization. Bob places a polarizing beam splitter (PBS) at the entrance of his interferometer, and rotates the polarization in one arm of the interferometer, for example by using a half wave plate (HWP) to cause the desired interference.}
\end{figure}

As a slight variation of the protocol, consider Alice encoding her outgoing pulses in different polarization, and Bob replacing his first beamsplitter by a polarizing beamsplitter (see Fig. \ref{pbs_fig}). This causes the two pulses to arrive simultaneously at Bob's second (interfering) beamsplitter. If he also rotates the polarization of the signal in one arm, all signals will interfere. 
We analyze the security proof of this protocol as well. Throughout the paper we will call this protocol the PBS protocol. 

\subsection{Proof techniques}

\begin{figure} 
\centering
\includegraphics[width=\columnwidth]{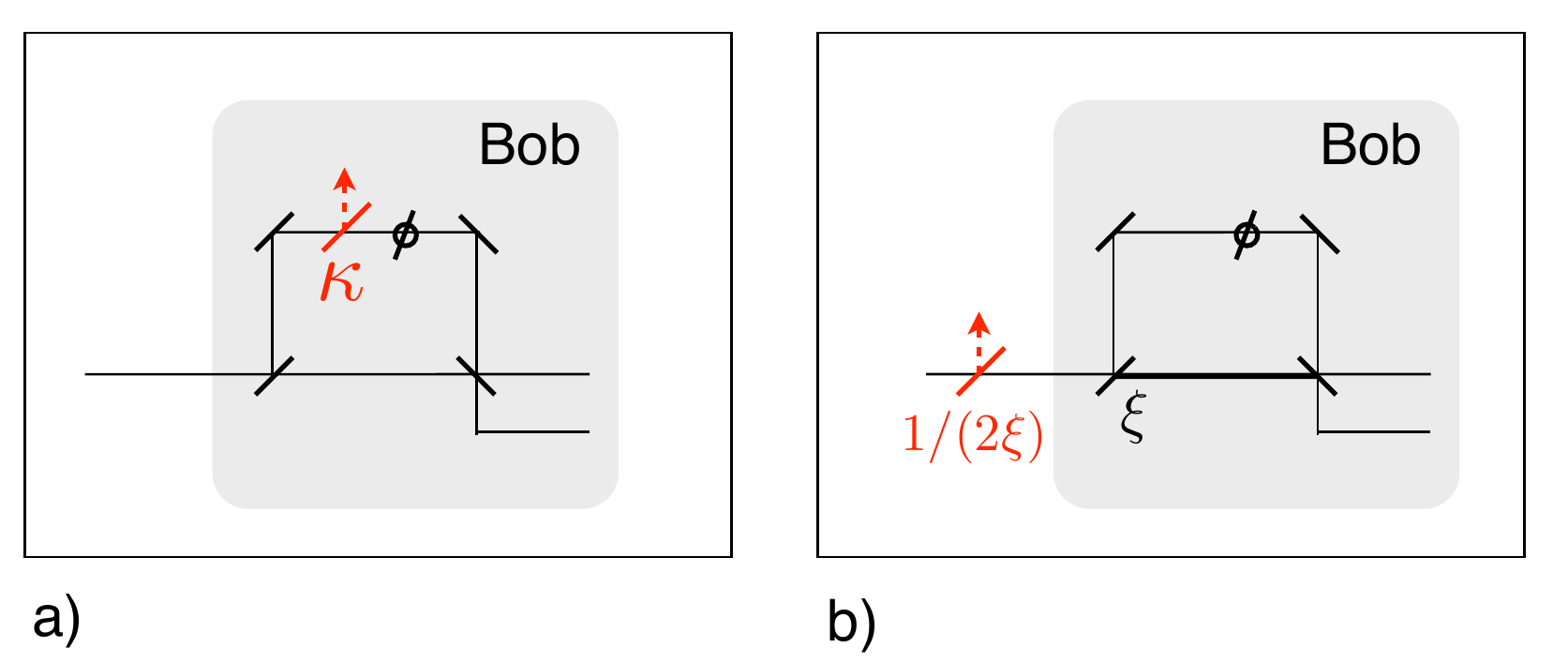}
\caption{ \label{loss-in-channel_fig} a) Original picture of the unbalanced phase-encoded protocol with the loss in Bob's interferometer. b) Equivalent lossless interferometer picture with a loss $1/(2\xi)$ in the channel followed by lossless interferometer with an uneven beam splitter with transmissivity $\xi$. }
\end{figure}

Typically, we assume that the eavesdropper (Eve) intercepts the quantum channel between Alice and Bob with the goal to learn about the key. Eve can tamper with the signals at will, but her interaction will introduce some disturbance to the signals, which can be detected by Alice and Bob.

It is generally difficult to provide a security proof for a scenario with a lossy measurement.  Therefore, we construct a picture where a lossy beamsplitter with transmissivity $\frac{1}{2\xi}$ is placed into the quantum channel followed by a lossless interferometer with an uneven first beamsplitter with transmissivity 
\begin{align}
\xi = \frac{1}{1+\kappa} \label{xi}.
\end{align}
This ``lossless interferometer'' picture is equivalent to the original picture with the lossy phase modulator in the long arm of the interferometer, as it yields the same measurement outcomes. The two pictures are shown in Fig. \ref{loss-in-channel_fig}.
However, in the lossless interferometer picture we can deal with the loss (that has now been outsourced to the channel) by giving Eve control over it and treat it like regular channel loss in the security proof.

In the case of the PBS protocol, we allot the control over the polarizing beam splitter and the lossy beamsplitter with transmissivity $\kappa$ in the long arm to Eve, leaving Bob with a lossless detector.


\section{Hardware fix}

\begin{figure} 
\centering
\includegraphics[width=\columnwidth]{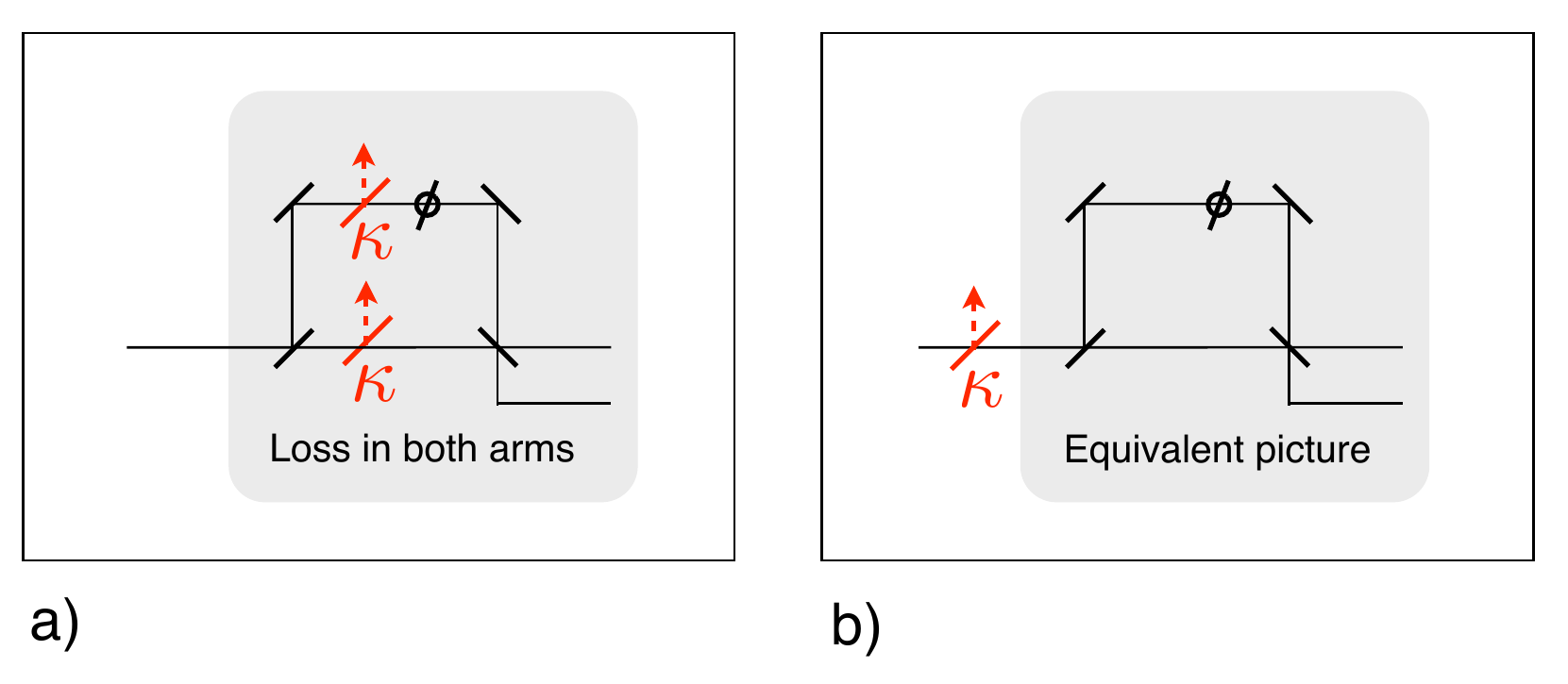}
\caption{ \label{hardwarefix-loss_fig} a) Hardware fix with the same amount of loss introduced in the short arm of the interferometer to compensate for the loss due to the phase modulator. b) Equivalent lossless interferometer picture.}
\end{figure}

\begin{figure} 
\centering
\includegraphics[width=\columnwidth]{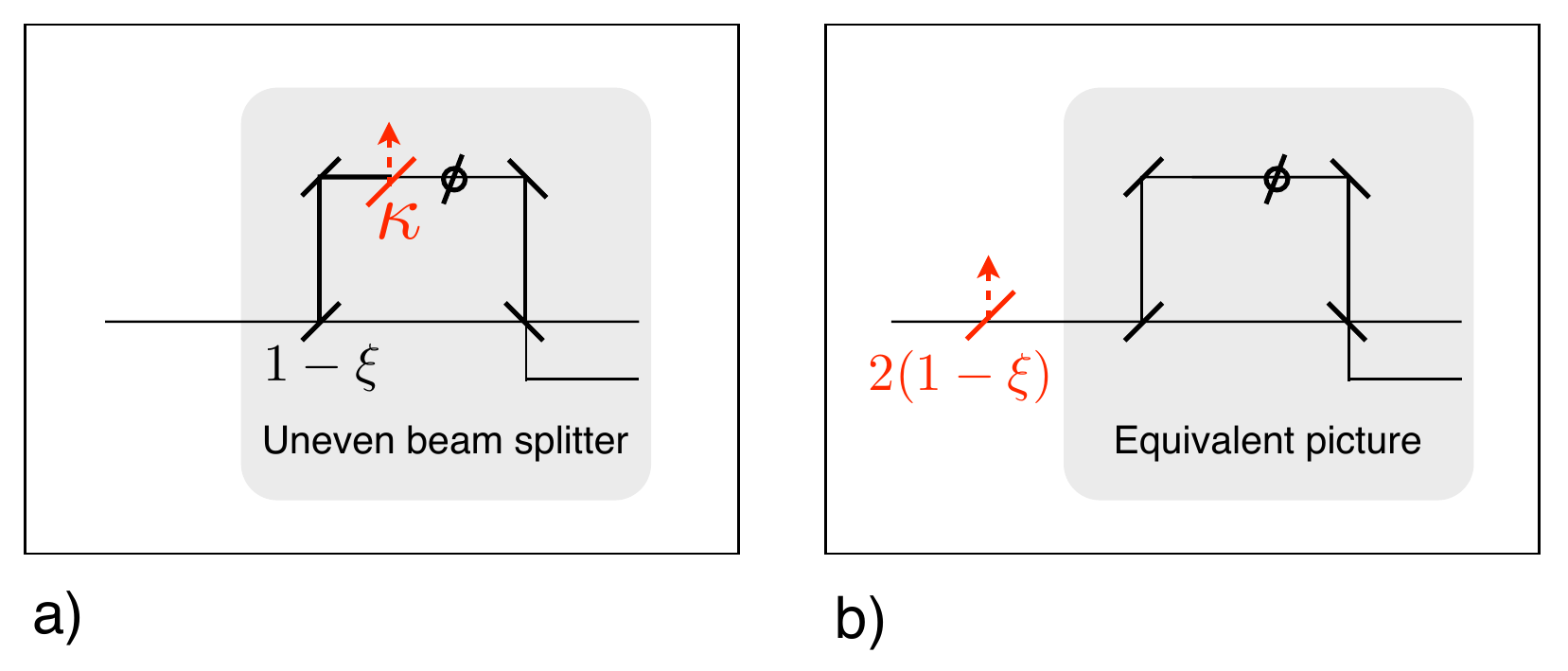}
\caption{ \label{hardwarefix-uneven_fig} Hardware fix with a biased beamsplitter in the interferometer to compensate for the loss due to the phase modulator. b) Equivalent lossless interferometer picture. }
\end{figure}

One simple way to recover the original BB84 scenario is by manually introducing a beamsplitter with the same transmissivity $\kappa$ in the shorter arm of the interferometers to compensate for the loss due to the phase modulator. Alternatively, one can replace the first beamsplitter in the interferometer by a biased beamsplitter with transmissivity $1-\xi$.
A schematic of these alternatives is shown in Figs. \ref{hardwarefix-loss_fig} and \ref{hardwarefix-uneven_fig}. 

The BB84 signal states and measurements are recovered in the equivalent lossless interferometer pictures, which are shown in Figs. \ref{hardwarefix-loss_fig} and \ref{hardwarefix-uneven_fig} for the two hardware fix possibilities. Under the assumption that the loss is attributed to Eve, the security proof reduces to the known BB84 security proof. 



\section{Single photon contribution}

In the following, we study the case where Alice sends a single photon and Bob obtains a single photon (qubit-to-qubit scenario). We analyze the signal structure and the measurements of the unbalanced phase-encoded protocol and the PBS protocol and provide a security proof (qubit security proof). Later, in section \ref{realistic_devices_sec}, we embed the qubit security proof into the more realistic scenario with optical modes in infinite dimensional Hilbert spaces using decoy states \cite{lo05a}, tagging \cite{gottesman04a, inamori07a} and squashing \cite{beaudry08a, tsurumaru08a, narasimhachar11asub}.

\subsection{Alice's signal states}

\begin{figure} 
\includegraphics[width=\columnwidth]{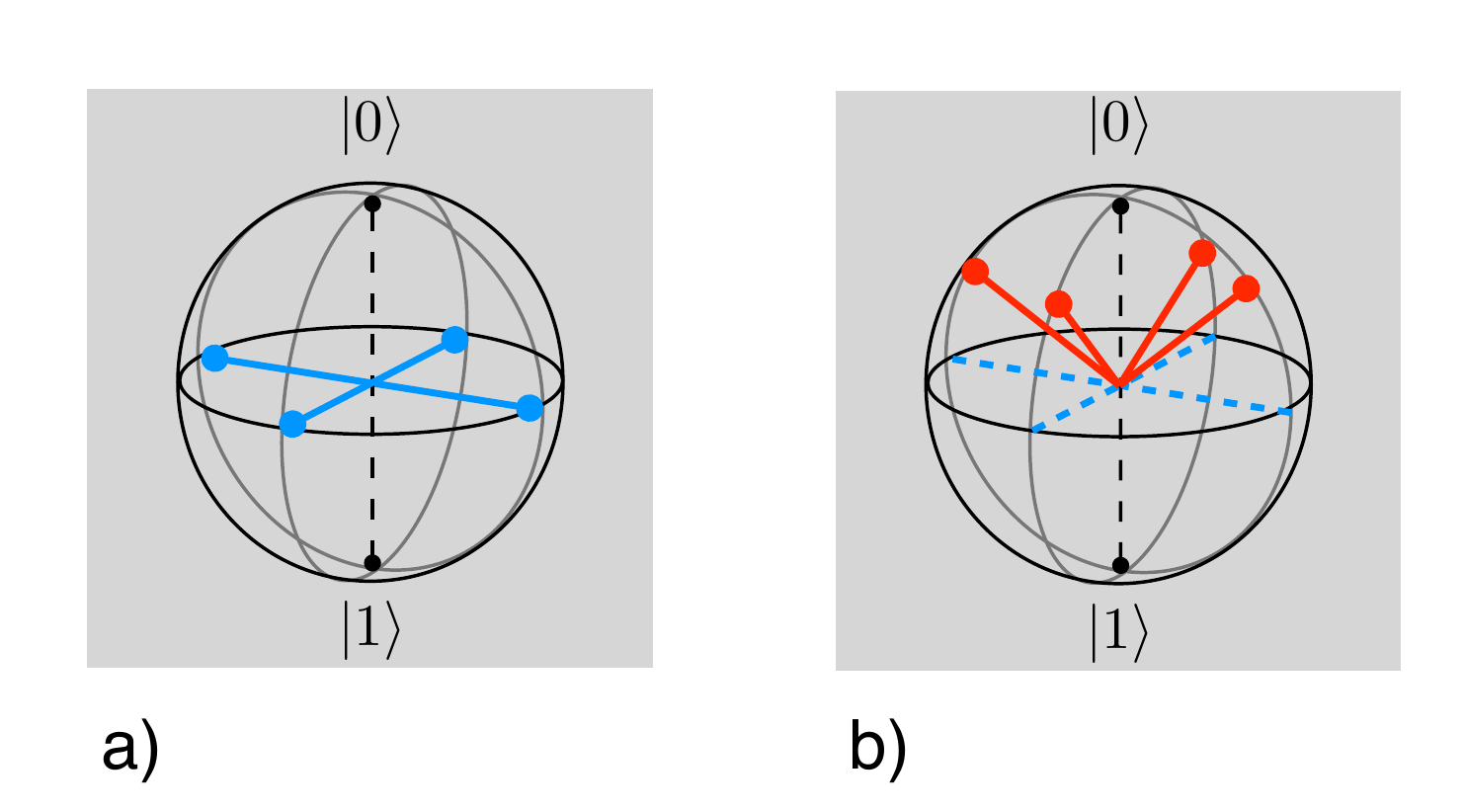}
\caption{\label{signal-states_fig} Comparison of the qubit signal states on the Bloch sphere. a) The signal states of the BB84 protocol, b) the signal states of the protocol with a lossy phase modulator.}
\end{figure}

If a single photon is distributed over the two time modes (pulses) emerging from Alice, the resulting Hilbert space of the signal states ($\mathcal H_S$) is a qubit space. 
We denote the creation operators of the two time modes by $a_0^\dagger$ and $a_1^\dagger$,  and define the two canonical basis vectors $\ket{0}= a_0^\dagger \ket{\mathrm{vac}}$ and $\ket{1} = a_1^\dagger \ket{\mathrm{vac}}$ of $\mathcal H_S$.
After Alice imprinted her phase choice onto the pulses, the signal leaving her apparatus can be in either of the four possible states
\begin{align}
\ket{\varphi_x} = \sqrt{\xi} \ket{0} + \sqrt{1-\xi} e^{ i \pi x/2}\ket{1}.
\end{align}
for $x  \in \{0, 1, 2 ,3 \}$.
The relationship between $\xi$ and $\kappa$ was already defined in Eq. (\ref{xi}).
In Fig. \ref{signal-states_fig} we show a representation of the signal state on the Bloch sphere in comparison to the signal states of the BB84 protocol. 

Alice then sends the signal states with equal probability through a quantum channel to Bob. This scheme is called prepare-and-measure scheme. Alternatively, the distribution of the signal states is captured in the source-replacement scheme: 
Alice constructs a (hypothetical) entangled state $\ket{\Phi} \in \mathcal H_{AS}$ in her lab, and sends the second system ($S$) to Bob. By means of a positive operator valued measure (POVM) Alice performs a measurement on the system $A$, which effectively prepares the signal states on the system $S$ for Bob. Furthermore, the reduced density matrix $\rho_A = \tr_S (\ketbra{\Phi}{\Phi})$ is fixed in the source-replacement scheme. In our case, the entangled source state is 
$\ket{\Phi}_{AS}  = \sqrt{\xi} \ket{00} + \sqrt{1-\xi} \ket{11}$
with a reduced density matrix 
\begin{align}
\rho_A = \xi \ketbra{0}{0} + (1-\xi) \ketbra{1}{1} \label{rhoA}.
\end{align}
Alice's POVM elements on $\mathcal H_{A}$ are then essentially BB84 measurements
\begin{align}
A_x = \frac{1}{2} P \bigg [ \frac{\ket{0} + e^{-i \pi x/2} \ket{1}}{\sqrt{2}}  \bigg ]. \label{Ax}
\end{align}
for $x \in \{0, 1, 2, 3 \}$. We denote by $P \big [\ket{\alpha} \big ] = \ketbra{\alpha}{\alpha}$ a projector.

\subsection{Bob's detection in the unbalanced phase-encoded protocol}

The modes $a_0^\dagger$ and $a_1^\dagger$ arrive at Bob's detector after Eve interacted with the signals. The output of Bob's detectors carry 6 modes in total, 2 in each of the 3 time slots. We denote the modes of the top (bottom) detector by $c_i^\dagger$ ($d_i^\dagger$) for $i=1,2,3$, respectively. For a fixed phase $\varphi_B \in \{ 0, \pi/2 \}$, the transformation of the $a_i^\dagger$ to the modes in the top and bottom detectors yield (up to global phases): 
\begin{align}
&c_1^\dagger= \frac{1}{\sqrt{2}} \sqrt{\xi} a_0^\dagger,  \nonumber \\
&c_2^\dagger= \frac{1}{\sqrt 2} ( \sqrt{1-\xi} a_0^\dagger + e^{i \varphi_B} \sqrt{\xi}  a_1^\dagger),  \nonumber \\
&c_3^\dagger= \frac{1}{\sqrt{2}} \sqrt{1-\xi} a_1^\dagger, \nonumber \\
&d_1^\dagger= \frac{1}{\sqrt{2}} \sqrt{\xi} a_0^\dagger,  \nonumber \\
&d_2^\dagger= \frac{1}{\sqrt 2} (  \sqrt{1-\xi} a_0^\dagger - e^{i\varphi_B} \sqrt{\xi}  a_1^\dagger), \nonumber \\
&d_3^\dagger= \frac{1}{\sqrt{2}} \sqrt{1-\xi} a_1^\dagger.  \nonumber
\end{align}
We choose to combine several outputs into one POVM element of Bob's measurement (coarse-graining). In terms of the incoming modes $\ket{0}$ and $\ket{1}$, Bob's POVM elements are 
\begin{align}
&B_y = \frac{1}{4} P[  \sqrt{1-\xi} \ket{0} + \sqrt{\xi} e^{ i \pi y/2} \ket{1})],\nonumber \\
&B_{\mathrm{out}} = \xi \ketbra{0}{0} + (1- \xi) \ketbra{1}{1}, \label{By}
\end{align}
where for $y$ runs over $\{0, 1, 2, 3 \}$. The 4 POVM elements $B_y$ correspond to inside clicks (time slot 2) in the two bases, while $B_{\mathrm{out}}$ denotes the POVM element of the outside clicks (time slots 1 and 3). 
 
\subsection{Bob's detection in the PBS protocol}

In a similar manner, the modes in the output ports of Bob's interferometer in the PBS protocol are found to be
\begin{align}
&c^\dagger = \frac{1}{\sqrt 2} (e^{-i\varphi_B} a_0^\dagger - i a_1^\dagger)  \nonumber,\\
&d^\dagger = \frac{1}{\sqrt 2} (-i e^{-i\varphi_B} a_0^\dagger + a_1^\dagger)  \nonumber.
\end{align}
There are only two output modes, since there are no outside clicks in this protocol. 
Bob's corresponding POVM on the input system has 4 elements for $y \in \{0, 1, 2, 3 \}$
\begin{align}
B'_y = \frac{1}{2} P \bigg [ \frac{(\ket{0} + e^{i \pi y /2} \ket{1}}{\sqrt{2}}  \bigg ], \label{ByPBS}
\end{align}
which is essentially a BB84 measurements. 

Indeed, in the PBS scenario, both Alice's and Bob's measurements are BB84 measurements. The difference between the PBS protocol and the BB84 protocol, however, is found in the reduced density matrix $\rho_A$, which contains the information about the modified signal structure as shown in Fig. \ref{signal-states_fig}.


\section{Description of the optimal attack}

Eve's interaction with the signals typically introduces some disturbance to the source state $\ket{\Phi}$. The actual state that Alice and Bob share after Eve's interaction is a general unknown (mixed) state $\rh$. 
The only knowledge Alice and Bob hold about $\rh$ is concentrated in the probability distribution
\begin{equation}
p(x,y)=\tr \{ A_x \otimes B_y \; \rho_{AB} \} \label{pxy}
\end{equation}
of their measurement outcomes with respect to the POVMs $\MA = \{ A_x \}$ and $ \MB =  \{ B_y \}$. These probabilities are determined during the step of parameter estimation. 
Additionally, they also know that Alice's reduced density matrix $\rho_A = \tr_B(\rh)$ is fixed, because it never entered Eve's domain. Consequently, they hold a parametrization of the set of all possible density operators $\rh$ that are compatible with $p(x,y)$ and $\rho_A$, which we define in the following by $\Gamma$:
\begin{definition} \label{def_Gamma}
The set $ \Gamma$ contains all bipartite states $\rh$ that have a given reduced state $\rho_A$ and are compatible with the measurement outcomes $p(x,y)$.
\end{definition}

In the source-replacement scheme Eve's attack is describe by the purification $\ket{\Psi}_{ABE}$ of $\rh$. The purification lives on a dilated space $\mathcal H_{ABE}$, where the dimension of the purifying system $E$ is the same as the dimension of $AB$. Giving  Eve full control over $\ket \Psi_{ABE}$, allows her to exploit everything allowed by quantum mechanics for her attack.


\section{Key rate formalism}

Along the lines of Ref. \cite{ferenczi12a},  we derive the key rate that Alice and Bob can extract from a state $\ket{\Psi}$ using postselection, such as basis sifting.

\subsection{Key rate formula}

Given Alice, Bob and Eve share the state $\ket{\Psi}$. 
After the measurement with respect to the POVMs $\MA = \{ A_x \}$ and $\MB = \{B_y\}$, the three parties share a classical-classical-quantum (ccq) state \cite{devetak05a}
\begin{eqnarray}
\rho_{XYE} = \sum_{x,y \in \mathbf M} p(x,y) \ketbra{x,y}{x,y} \otimes \rho_{E}^{xy}.
\end{eqnarray}
The probability $p(x,y)$ was already defined in Eq. (\ref{pxy}), whereas Eve's conditional states are given by $\rho_E^{xy} = \tr_{AB}\{ A_x \otimes B_y \otimes \mathbbm 1_E \ketbra{\Psi}{\Psi} \}/p(x,y)$. 
According to Refs. \cite{devetak05a, kraus05a}, the key rate that can be extracted from such a state is
\begin{align}
&r(\rho_{XYE}) =  I(X:Y) - \chi(X:E),
\end{align}
where $I(X:Y)=H(X)+H(Y)-H(X,Y)$ is the classical mutual information between Alice and Bob's data $p(x,y)$, and $\chi (X:E) = H(X) + S(E) - S(X,E)$ is the Holevo quantity between Alice and Eve. $H$ and $S$ denote the Shannon entropy and the von Neumann entropy, respectively. If $\rh$ is known, the Holevo quantity can be expressed in terms of Eve's states $\rho_{E}^{x}$ conditioned on Alice's value $x$ and her reduced state $\rho_{E}=\sum_{x} p(x) \rho_{E}^{x}$ 
\begin{eqnarray}
\chi (X:E) =S(\rho_{E})-\sum_{x} p(x) S(\rho_{E}^{x})  \label{Holevo}.
\end{eqnarray} 
If Alice's POVM elements are rank-one projectors, an explicit reference in the Holevo quantity to the system $E$ can be eliminated. It can be expressed in terms of quantities on the systems $AB$ only:
\begin{eqnarray}
\chi (X:E) =S(\rho_{AB})-\sum_{x} p(x) S(\rho_{B}^{x})  \label{HolevoAB}.
\end{eqnarray}

\subsection{Key rate formula with postselection}

First we introduce the same notation as in Ref. \cite{ferenczi12a} with emphasis on the dependence of the key rate on $\rho_{AB}$ and the corresponding measurement, rather than the explicit ccq state $\rho_{XYE}$. In the new notation we denote the Holevo quantity by 
\begin{align}
&\chi(\rh, \MA) :=\chi(X:E).
\end{align}

In the classical description of the postselection, Alice and Bob make an announcement to each measurement outcome to filter out uncorrelated data.
Alice announces the basis of her outcome (``even" or ``odd"), and Bob announces the basis of his outcome (``even" or ``odd") if it was an inside click. If it was an outside click, he and announces ``out". These announcements do not reveal the bit values. Based on the announcements, they keep the events where both announced ``even" or both announced ``odd", and discard the rest. We denote the events where they had the same announcement by $u = $ ``even'' or $u = $ ``odd''.
The identification of the announcements effectively defines the subsets $\mathbf m_A^\even = \{ A_0, A_2 \} $, $\mathbf m_A^\odd = \{ A_1, A_3 \} $, $\mathbf m_{B}^\even = \{ B_0, B_2 \} $ and $\mathbf m_{B}^\odd = \{ B_1, B_3 \} $ of the POVMs $\mathbf M_A$ and $\mathbf M_B$, which contain the POVM elements corresponding to the announcement $u = $ ``even'' or $u = $ ``odd''. 

The quantum version of the postselection is described by a two-step process on $\rho_{AB}$: first, a trace-preserving quantum map $\mathcal E$ is applied to $\rho_{AB}$ that takes care of the announcement and discarding. The Kraus operators of the map are given by $F_A^u \otimes F_B^u$ corresponding to the coinciding announcements $u$. The filters are explicitly given by $F_A^u = \sqrt{\sum_{\mathbf m_A^u}A_x}$ and $F_B^u=\sqrt{\sum_{\mathbf m_B^u}B_y}$.

This map creates an ensemble $\{\Fu[\rh], p(u)\}$ of normalized states 
\begin{align}
& \Fu[\rh] = F_A^u \otimes F_B^u \rho_{AB} (F_A^u \otimes F_B^u)^\dagger / \tilde p(u) \label{rhoABu},\\
& \tilde p(u) = \tr \{ F_A^u \otimes F_B^u \rho_{AB} (F_A^u \otimes F_B^u)^\dagger  \} \label{tildepu},
\end{align}
each with probability $p(u)= \frac{\tilde p(u)}{p_{\mathrm{kept}}}$. The probability that $\rh$ is kept in this process is $p_{\mathrm{kept}} = \sum_u \tilde p(u)$. 

In the second step, the states $\Fu[\rh]$ are measured with respect to new measurements $\MAu \otimes \MBu$ conditioned on $u$. The new POVMs $\MAu$ and $\MBu$ are found by renormalizing the sets $\mathbf m_A^u$ and $\mathbf m_B^u$ by the pseudo-inverses of the filters defined on the non-zero subspace of $F_A^u \otimes F_B^u$. This ensures that measuring $\Fu[\rh]$ with respect to $\MAu \otimes \MBu$  yields the same outcomes as measuring $\rho_{AB}$ with respect to $\MA \otimes \MB$:
\begin{align}
&\mathbf M_A^u  = \{(F_A^{u})^{-1} A_x (F_A^{u \dagger})^{-1} : A_x \in \mathbf m_A^u \}  \nonumber, \\
&\mathbf M_B^u  = \{ (F_B^{u})^{-1} B_y (F_B^{u \dagger})^{-1} : B_y \in \mathbf m_B^u \}.
\end{align}

In \cite{ferenczi12a}, the key rate after post-selection ($\bar r$) extracted from each coinciding announcement $u$ independently is given by
\begin{align}
\bar r(\mathcal E (\rho_{AB})) = \bar I_{\mathrm{obs}}  - \bar \chi(\mathcal E (\rho_{AB})),
\end{align}
with the following definition of the overall Holevo quantity
\begin{align}
\bar \chi(\mathcal E (\rho_{AB}))  :=  \sum_u p(u) \chi(\Fu[\rh], \MAu) \label{barchi}.
\end{align}
The overall mutual information $\bar I_{\mathrm{obs}}$ is the mutual information of Alice and Bob's data after postselection. It is fixed by the measurement outcomes $p(x,y)$.

\subsection{Description of Eve's optimal attack}

Typically, the state $\ket{\Psi}$ is not completely know to Alice and Bob. They might hold only partial information about $\ket{\Psi}$ via the characterization of the set $\Gamma$  in the parameter estimation step.
If this is the case, Alice and Bob must assume that Eve chose the most powerful attack (optimal attack) compatible with $\Gamma$, which is the attack that yields the lowest key rate
\begin{align}
r_{\mathrm{min}} & = \min_{\rho_{AB} \in \Gamma} \bar r(\mathcal E (\rho_{AB}))  \nonumber \\
&= \bar I_{\mathrm{obs}} -  \max_{\rho_{AB} \in \Gamma} \bar \chi(\mathcal E (\rho_{AB})) \label{rmin}.
\end{align}
While the first term ($\bar I_{\mathrm{obs}}$) is fixed by Alice and Bob's measurement outcomes $p(x,y)$, the second term ($\bar \chi$) is obtained through optimization over all possible attacks.


\section{Symmetries in protocols}

In this section we describe a scenario in which we can show that the optimal attack carries a certain symmetry without loss of generality. Under the assumption that Alice and Bob use a coarse-grained version of the exact probability distribution $p(x,y)$ for the parameter estimation, and if the overall Holevo quantity $\bar \chi$ satisfies certain symmetry properties, it can be shown that the optimal attack lies within a symmetric set. The symmetry property of the optimal attack is useful in the optimization of the key rate in Eq. (\ref{rmin}). For a detailed analysis see Ref. \cite{ferenczi12a}. 

First, let us define the symmetries of the POVMs $\MA$ and $\MB$.
Let $G$ be a symmetry group with group elements $g$ and a unitary representation $U_g$ on the Hilbert space $\mathcal H_S$. A set $\mathbf S$ containing operators on $\mathcal H_S$ is called $G$-invariant, if for all elements $s_x \in \mathbf S$ and all $U_g \in G$ it holds that $U_g s_x U_g^\dagger :=s_{g(x)} \in \mathbf S$.

Suppose our protocol is equipped with POVMs $\MA$ and $\MB$ that exhibit the following $G$-invariance
\begin{align}
&U_g^* A_x U_g^T =: A_{g(x)} \in \MA \label{Axinv},\\
&U_g B_y U_g^\dagger =: B_{g(y)} \in \MB \label{Byinv}.
\end{align}
Consider now a state 
\begin{align}
\Ug[\rh] = U_g^* \otimes U_g \rho_{AB} U_g^T \otimes U_g^\dagger \label{rhoABUg},
\end{align}
which is unitarily equivalent under the symmetry group to $\rh$. If we measure $\Ug[\rh]$ with respect to $\MA \otimes \MB$, the resulting probability distribution, $p_g(x,y)$, of the measurement outcomes is, by virtue of the $G$-invariance, a permuted version of $p(x,y)$. Furthermore, if $\rho_A = U_g^* \rho_A U_g$, the reduced density matrix of $\Ug[\rh]$ is unchanged. 

In the parameter estimation Alice and Bob are free to use a coarse-grained version $Q[\{p(x,y)\}]$ of the detailed probability distribution $p(x,y)$. In fact, in many cases, only averaged observations, like the average quantum bit error rate, are used for this task. The averaged observations that we choose to use in our protocol are invariant under permutation of the $p(x,y)$, in the sense that
\begin{align}
&Q[\{ p(x,y)\} ] = Q [\{ p_g(x,y) \}] \label{Qinvariance}.
\end{align}
In this scenario, all states $\Ug[\rh]$ are compatible with $Q$, whenever $\rh$ was compatible with $Q$. 
If $Q$ is also linear in $p(x,y)$, we can also include the convex combination
\begin{align}
\barrh = \frac{1}{|G|} \sum_{g \in G}  \; \Ug[\rh] \label{barrhoAB}
\end{align}
to the set of compatible states. Here, $|G|$ is the cardinality of the group $G$. 
Consequently, we define the new set $\Gamma_{\mathrm{ave}}$, containing all density operators which are compatible with $Q$ and $\rho_A$. The new set $\Gamma_{\mathrm{ave}}$ is a superset of the original set $\Gamma$. 

As shown in \cite{ferenczi12a}, if the Holevo quantity is concave in $\rh$ and satisfies the equivalence property,
$ \bar \chi(\mathcal E ( \rho_{AB})) =  \bar \chi(\mathcal E (\Ug[\rh]))$, the quantity $\bar \chi(\mathcal E (\barrh))$ calculated from the symmetrized state $\barrh$ always upper bounds the quantity $\bar \chi(\mathcal E (\rh))$ calculated from $\rh \in \Gamma_{\mathrm{ave}}$. Therefore, without loss of generality, the optimal attack is found in a set $\bar \Gamma$ defined as follows: 
\begin{definition} \label{def_Gamma}
The set $\bar  \Gamma$ contains all bipartite symmetrized states $\barrh$ that are compatible with $Q$ and that have a given reduced state $\rho_A$.
\end{definition}
We then obtain the key rate:
\begin{align}
r_{\mathrm{min}} &= \bar I -  \max_{\bar \rho_{AB} \in \bar \Gamma} \bar \chi(\mathcal E (\bar \rho_{AB})).
\end{align}


\section{Convexity and equivalence property of $\bar \chi$ in the example of the unbalanced phase-encoded protocol}

In this section we show the concavity and equivalence property of $\bar \chi$ in the case of the unbalanced phase-encoded protocol. 
We then compute the key rate in the scenario with the coarse-grained parameter estimation.

\subsection{Postselection}

In the sifting process Alice and Bob postselect on events in the same basis (``odd'' or ``even'') in the middle time slot. We first calculate Alice's filters for this postselection from the ``odd'' and ``even'' POVM elements in Eq. (\ref{Ax}):
\begin{align}
&F_A := F_A^\odd = F_A^\even =  \mathbbm 1 /\sqrt 2 \label{FKequA}.
\end{align}
The new renormalized POMV measurements $\MAu$ conditioned on $u$ are related to the original POVM elements in $\mathbf m_A^{u}$ simply by a factor 2:
\begin{align}
&\mathbf M_A^\even  = \{2 A_0, 2 A_2 \}  \nonumber, \\
&\mathbf M_A^\odd  = \{2 A_1, 2 A_3 \} .
\end{align}
We also calculate Bob's filters from the ``odd'' and ``even'' POVM elements in Eq. (\ref{By}). There is a dependence on $\xi$ in Bob's filter, but again the filters are equal for ``even'' and ``odd'' bases
\begin{eqnarray} 
F_{B}:= F_{B}^{\mathrm{even}} = F_{B}^{\mathrm{odd}} = \frac{1}{\sqrt{2}} \label{FKequB}
\left( \begin{array}{cc} 
\sqrt{1-\xi} & 0  \\
0 & \sqrt{\xi}
\end{array} \right).
\end{eqnarray}
Here the filters are presented in the basis $\{\ket 0, \ket 1 \}$.
It is straightforward to find Bob's post-selected POVMs 
\begin{align}
&\mathbf M_{B}^\even  = \{2 B'_0, 2 B'_2 \},  \nonumber \\
&\mathbf M_{B}^\odd = \{2 B'_1, 2 B'_3 \} ,
\end{align}
expressed in terms of the BB84-type measurements $B'_y$ in Eq. (\ref{ByPBS}).

\subsection{Symmetries}

The symmetry group $G$ governing the POVMs $A_x$ and $B_y$ is the cyclic group $C_4$ with 4 elements. A reducible representation of $C_4$ in the canonical basis of the signal states is given by
\begin{align}
U_g = \left( \begin{array}{cc} 
1 & 0  \\
0 & e^{i g \pi/2 }
\end{array} \right)  \quad g \in \{0,1,2,3 \}.
\end{align}
Alice and Bob's POVMs satisfy the symmetry relation $ U_g^* A_x U_g^T = A_{x+g}$ and $U_g B_y U_g^\dagger = B_{y+g}$  where addition is taken modulo 4, while the outside element $B_{\mathrm{out}}$ and the reduced density matrix $\rho_A$ in Eq. (\ref{rhoA}) remain invariant under the action of $U_g$.   
Note that complex conjugation and Hermitian conjugation are equivalent operations for all unitaries $U_g$, because they are simultaneously diagonal.

Not only are the POVMs $\MA$ and $\MB$ $C_4$-invariant, but also the post-selected POVMs satisfy a certain symmetry relation with respect to $C_4$. 
Let us define the action of a unitary $U$ on a POVM $\mathbf M = \{K_x\}$ by 
\begin{align}
&U \mathbf M U^\dagger := \{U K_x U^\dagger \}
\end{align}
The $C_4$-invariance of the sets of POVMs $\MAu$ and $\MBu$ is given by 
\begin{align}
U_g^* \MAu U_g^T = \mathbf M_A^{g(u)}   \nonumber, \\
U_g \MBu U_g^\dagger = \mathbf M_B^{g(u)} \label{GinvMu}.
\end{align}
For this particular example, $g(u) = u \oplus \mathrm{parity}(g)$ with $\mathrm{parity}(g) \in \{\even, \odd \}$. The addition is defined by the rules: $\even \oplus \even = \odd \oplus \odd = \even$ and $\odd \oplus \even = \even \oplus \odd = \odd$.

\subsection{Concavity and equivalence property of $\bar \chi$}

From the property that the filters are equal for ``even'' and ``odd'' announcements in Eqs. (\ref{FKequA}, \ref{FKequB}), 
we can derive the relationship between ``even'' and ``odd'' normalization $\tilde p(\odd) = \tilde p(\even) =: \tilde p$ and the postselected density matrices:
\begin{align}
& \F[\rh] := \mathcal F^\even[\rh] = \mathcal F^\odd[\rh] \label{equrho}.
\end{align}
While $\tilde p(u)$ depends on the density matrix $\rho_{AB}$, the normalized probability  distribution $p(u)$ does not: $p(\odd)=p(\even)= \frac{1}{2}$.
Furthermore, the filters commute with all unitaries $U_g$ for Alice and Bob:
\begin{align}
[F_K,U_g] =0 \quad \forall g\in G, \; K \in \{A,B \} \label{FKcomm}.
\end{align}
Consequently, $\F$ in Eq. (\ref{equrho}) commutes with the symmetry group
\begin{align}
\F[\Ug[\rh]] = \Ug[\F[\rh]] \label{sigu}.
\end{align}
Moreover, $\F$ acts linearly on any convex combination of states of the form $\Ug[\rh]$, for example on the symmetrized state $\barrh$
\begin{align}
\F[\barrh] = \frac{1}{|G|} \sum_{g} \F [ \Ug[ \rh]]  \label{barrhou}.
\end{align}

We have now all ingredients to show the concavity and the equivalence property of $\bar \chi$. 
Consider three density matrices $\rh$, $\sig$ and the convex combination $\barrh = \lambda \rh + (1-\lambda) \sig$ with $\lambda \in [0,1]$.  
In Ref. \cite{ferenczi12a} it is shown that the quantity $\chi(\rho_{AB}, \MA)$ is concave with respect to $\rh$, namely:
\begin{align}
\chi(\bar \rho_{AB}, \MA) \geq \lambda \chi(\rho_{AB}, \MA) + (1-\lambda) \chi(\sigma_{AB}, \MA) \label{concav}.
\end{align}
Using Eq. (\ref{barrhou}) and (\ref{concav}), the concavity of $\bar \chi$ follows:
\begin{align}
\bar \chi(\mathcal E (\bar \rho_{AB})) &= \frac{1}{2} \sum_u \;  \chi \left (  \F[\barrh], \MAu \right ) \nonumber \\
&\geq \frac{1}{2} \sum_u  \frac{1}{|G|} \sum_g \chi \left ( \F[\Ug[\rh]], \MAu  \right ) \\
&= \frac{1}{|G|} \sum_g  \bar \chi(\mathcal E ( \Ug[\rh]).  \nonumber
\end{align}

For the equivalence property we resort to lemma 1 in \cite{ferenczi12a} that describes the invariance property of the Holevo quantity under unitary transformation 
\begin{align}
\chi(\Ug[\rh], \MA) = \chi(\rh, U_g^T \MA U_g^* ).
\end{align}
This property also holds for the postselected states due to the commutation rule in Eq. (\ref{sigu}) and the $G$-invariance of the $\MAu$ in Eq. (\ref{GinvMu})
\begin{align}
\chi \left(\F[\Ug[\rh]],  \mathbf M_A^{g(u)} \right ) 
= \chi \left (\F[\rh], \MAu \right ). \label{chiinv} 
\end{align}
The equivalence transfers to $\bar \chi$ by means of Eqs. (\ref{chiinv}) and (\ref{equrho}):
\begin{align}
\bar \chi(\mathcal E (\rh))  &=  \frac{1}{2} \sum_u \chi(\F[\rh], \MAu) \nonumber \\
&= \frac{1}{2} \sum_u  \chi(\F[\Ug[\rh]],  \mathbf M_A^{g(u)})\\
&=\frac{1}{2}  \sum_{g(u)}  \chi(\F[\Ug[\rh]],  \mathbf M_A^{u}) = \bar \chi(\mathcal E (\Ug[\rh])) \nonumber.
\end{align}

Having showed concavity and equivalence, we can assume without loss of generality that the optimal attack is chosen from the set $\bar \Gamma$. The two main properties used to prove the equivalence and concavity were essentially Eqs. (\ref{FKequA}, \ref{FKequB}) and (\ref{FKcomm}).


\section{Key rates in the qubit-to-qubit scenario} \label{unbalanced_phase-encoded_qubit_sec}

We now calculate the key rate of the unbalanced phase-encoded protocol in the qubit-to-qubit scenario. We assume that Alice and Bob perform the parameter estimation based on the average error rate
\begin{align}
Q = \frac{p(0,2) + p(2,0) + p(1,3) + p(3,1)}{2 \tilde p} \label{Q},
\end{align}
which is calculated from the detailed probability distribution $p(x,y)$ and the normalization $\tilde p$ in Eq. (\ref{tildepu}). The average error rate satisfies the property (\ref{Qinvariance}), which follows from the $G$-invariance of $\MAu$ and $\MBu$ in Eq.(\ref{GinvMu}). 

The symmetric state $\barrh$ with respect to the $C_4$ symmetry group is generally described by the density matrix 
\begin{eqnarray} \label{barrhoABmat}
\bar \rho_{AB} = \left( \begin{array}{cccc} 
a & & & f^* \\
& b & & \\
& & c & \\
f & & & d  \end{array} \right) 
\end{eqnarray}
with open parameters $a$, $b$, $c$, $d$ and $f$ that satisfy the trace conditions $\tr{\barrh} = 1$. The matrix representation of $\barrh$ is with respect to the canonical basis $\{ \ket{ 00}, \ket{ 01} , \ket{ 10}, \ket{1 1} \}$.
The reduced density matrix 
\begin{align}
&\tr_B ( \barrh) = \left( \begin{array}{cc} 
a+b &  \\
& c+d   \end{array} \right) \label{rhoAparam}
\end{align}
is fixed by $\rho_A$ (see Eq. (\ref{rhoA})), which leads to the additional constraint $a+b = \xi$ on the parameters.
Furthermore, the error rate constraint 
\begin{align}
Q  = \frac{\tilde p - \frac{1}{2} \Re [f] \sqrt{\xi (1-\xi)}}{2 \tilde p}
\end{align}
eliminates another parameter. The normalization $\tilde p$ is given by
$\tilde p = \frac{1}{4}((1-\xi)(a+c) + \xi (b+d))$ and $\Re[f]$ denotes the real part of $f$. 

We calculate the Holevo quantity $\bar \chi$ and use MATLAB calling the optimization function fmincon to perform a numerical optimization over the states $\barrh \in \bar \Gamma$. 
To simulate the classical data used for the calculation of the mutual information, we assume a typical scenario with symmetric observations. The mutual information is then a function of the average error rate $\bar I_{\mathrm{obs}}  = 1-h(Q)$, with $h$ denoting the binary entropy. 

\begin{figure} 
\centering
\includegraphics[width=\columnwidth]{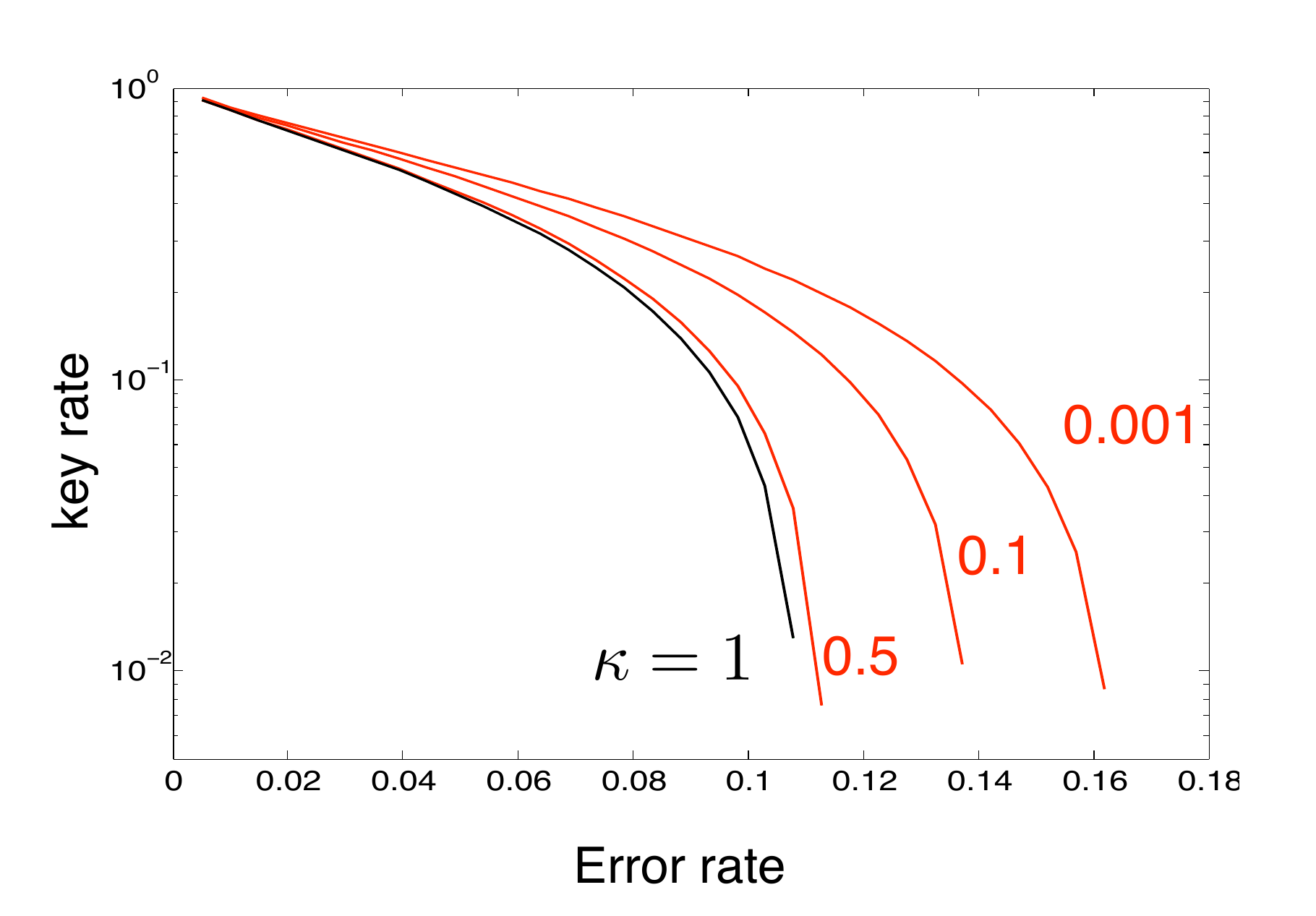}
\caption{\label{key-rate-Q} Key rates per postselected signal in the middle time slot for unbalanced phase-encoded protocol in dependence on the error rate $Q$ for different values of $\kappa$ in the phase modulator.}
\end{figure}

In Fig.\ref{key-rate-Q} we show a plot of the key rates renormalized on matching bases and clicks in the  middle time slot for different values of $\kappa$ in the case of a lossless channel. In particular, observe that as $\kappa$ increases, the key rate increases as well. This behavior originates from the signal state structure: as the signals become more non-orthogonal with increasing $\kappa$, the eavesdropper has more difficulty to distinguish them. This advantage disappears quickly, though, once loss is added to the channel. 
For a fair comparison of the protocol performances, however, we must analyze the protocols under a realistic channel model. 

\section{Concavity and equivalence in the case of the PBS protocol}

The PBS protocol differs from the unbalanced phase-encoded protocol only in the measurements on Bob's side. Bob's filters for the PBS protocol,
\begin{align}
&F_{B'}: = F_{B'}^{\mathrm{even}} = F_{B'}^{\mathrm{odd}} =  \mathbbm 1 /\sqrt 2
\end{align}
also satisfy the properties in Eqs. (\ref{FKequA}, \ref{FKequB}) and (\ref{FKcomm}). As these were the two properties needed to show concavity and equivalence of $\bar \chi$, we can again assume a symmetric optimal attack of the form (\ref{barrhoABmat}).


\section{Key rates for realistic devices} \label{realistic_devices_sec}

The security proof in section \ref{unbalanced_phase-encoded_qubit_sec} is tailored to the situation where Alice and Bob use perfect qubits (e.g. single photons). 
The actual experimental implementations, however, are performed with optical modes in an infinite dimensional Hilbert space. 
In particular, Alice's device can send vacuum and multi-photons into the channel, and Bob's detector can receive vacuum and multi-photons from the channel.  
In order to achieve a complete security proof, we need to include the deviation from the ideal qubit-to-qubit scenario.

Recently, several powerful tools \cite{gottesman04a, inamori07a,  lo05a, beaudry08a, tsurumaru08a, narasimhachar11asub} have been developed to bridge the gap between theory and experiment, with the aim to extend the validity of qubit-based security proofs to the more realistic scenario of optical modes. On Alice's side, the multi-photon components are taken care of by using decoy states \cite{lo05a}, supported by tagging \cite{gottesman04a, inamori07a}. 
This essentially permits to estimate the fraction of the single photon contributions in the data. 
On the other hand, the multi-photons entering Bob's detector are taken care of by the squashing method in Refs. \cite{beaudry08a, tsurumaru08a, narasimhachar11asub}.
If a squashing map exists for a certain measurement setup, then the detection pattern resulting from an arbitrary input state into Bob's detector can be interpreted as if it were coming from a single photon or a vacuum input. 

Note that this approach gives a provable secure key rate, however, higher key rates may be achievable with a refined analysis. The starting point for the refined analysis is the observation that a photon number splinting attack on multi-photon signals does not leak the signal content with certainty to Eve, even after the basis announcement. However, such an analysis will be rather involved. It is expected that for strong asymmetries in the balancing the refined analysis would improve the results significantly, as the scheme then takes a similar form as the strong reference pulse schemes analyzed in \cite{koashi04a, tamaki04a}.

\subsection{Key rate for unbalanced phase-encoded protocol with realistic devices}

\begin{figure} 
\includegraphics[width=\columnwidth]{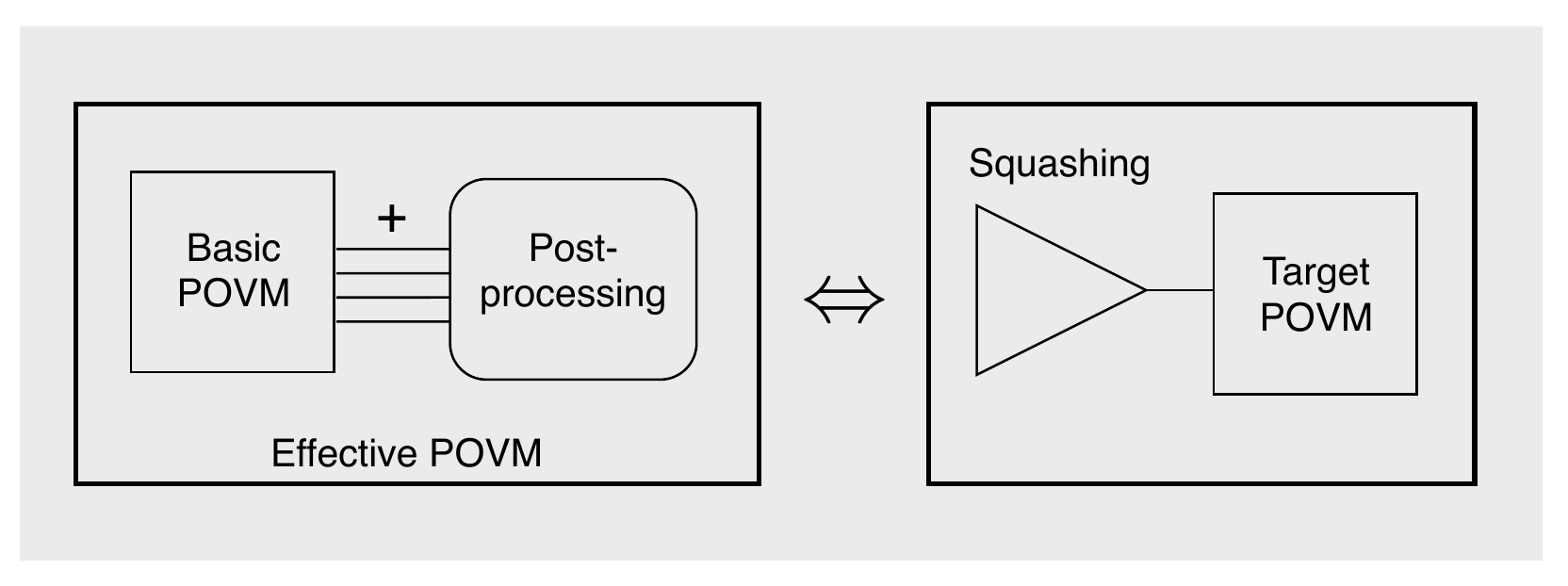}
\caption{\label{squashing_fig} The basic POVM combined with classical post-processing forms the effective POVM, which is mapped to the target (qubit) POVM by the squashing map.}
\end{figure}

For the unbalanced phase-encoded detector setup, a valid squashing map has been proven to exist in Ref. \cite{narasimhachar11asub}. A schematic of the squashing idea is shown in Fig. \ref{squashing_fig}.
In the proof it is assumed that Bob's detectors can resolve the three different time slots. 
Therefore, a general incoming optical mode can trigger any possible click pattern in the 6 detection slots. A POVM is associated to this measurement, called the basic POVM. 
Through classical post-processing, described in more detail below, several basic POVM elements are combined to form the effective (single-click) POVM elements.
The effective POVM has 5 elements $\{ B^{\mathrm{eff}}_0, B^{\mathrm{eff}}_1,  B^{\mathrm{eff}}_2,  B^{\mathrm{eff}}_3, B^{\mathrm{eff}}_{\mathrm{out}} \}$, which reflect the single-click structure of the target qubit POVMs given in Eq. (\ref{By}). The squashing map in Ref.  \cite{narasimhachar11asub} then maps the effective POVM to the target POVM. 

The post-processing in Ref. \cite{narasimhachar11asub} that maps the basic POVM to the effective POVM is shown in Table \ref{squashing_tab}. In order to maintain the statistics for an incoming single photon, any single-click basic POVM element is mapped to the corresponding effective POVM element. Furthermore, the basic multi-click POVM elements are processed as follows: 
double-click POVM elements in the ``even'' (``odd'') basis in the middle time slot are mapped with equal probability to $ B^{\mathrm{eff}}_0$ or $ B^{\mathrm{eff}}_2$ ($ B^{\mathrm{eff}}_1$ or $ B^{\mathrm{eff}}_3$); 
outside multi-click POVM elements are always mapped to the outside POVM element $B^{\mathrm{eff}}_{\mathrm{out}}$;  cross-click POVM elements (simultaneous middle and outside clicks) are probabilistically mapped with probability 0.5 to $B^{\mathrm{eff}}_{\mathrm{out}}$ and with probability 0.125 to each of the four $B^{\mathrm{eff}}_y$, $y \in \{0, 1, 2, 3 \}$. This post-processing essentially maps double-clicks to errors in the data.  

\begin{table} 
\caption{ Mapping of the basic POVM to the effective POVM via probabilistic classical post-processing.}\label{squashing_tab} 
\begin{tabular}{lcr}
\addlinespace[0.5em]
\toprule[0.1em]
\addlinespace[0.5em]
Basic 						& Effective 	& Probability \\
POVM 						& POVM 		& \\ \addlinespace[0.5em] \midrule[0.1em] \addlinespace[0.5em]
Single-click 					& Single-click 	& 1 \\  \addlinespace[0.2em] \hline \addlinespace[0.2em]
\multirow{2}{*}{Double-click (middle) ``even''} & $B^{\mathrm{eff}}_0$		&  0.5  \\ 
							& $B^{\mathrm{eff}}_2$  			&  0.5  \\ \addlinespace[0.2em] \hline \addlinespace[0.2em]
\multirow{2}{*}{Double-click (middle) ``odd''} & $B^{\mathrm{eff}}_1$		&  0.5  \\ 
							& $B^{\mathrm{eff}}_3$  			&  0.5  \\ \addlinespace[0.2em] \hline \addlinespace[0.2em]
Multi-click (outside only)				& $B^{\mathrm{eff}}_{\mathrm{out}}$ & 1\\ \addlinespace[0.2em] \hline \addlinespace[0.2em]
\multirow{5}{*}{Cross-click} 		& $B^{\mathrm{eff}}_0$ & 0.125 \\ 
							& $B^{\mathrm{eff}}_1$ & 0.125 \\ 
							& $B^{\mathrm{eff}}_2$ & 0.125 \\ 
							& $B^{\mathrm{eff}}_3$ & 0.125 \\ 
							& $B^{\mathrm{eff}}_{\mathrm{out}}$ & 0.5 \\ \addlinespace[0.5em]
\bottomrule[0.1em]
\end{tabular}
\end{table}

Due to the existence of the squashing map, we can assume that all detection events on Bob's side are single-photon events, whereas Alice's source can either send a vacuum signal ($v$), a single-photon ($s$) or multi-photons ($m$). 

In the error correction step, Alice and Bob must correct all errors in their data. Therefore, the error correction term, which is the mutual information, depends on the total observed error rate $Q_{\mathrm{tot}}$:  $\bar I_{\mathrm{}obs} = 1 - h(Q_{\mathrm{tot}})$. 
Due to the decoy states and the tagging method, however, the privacy amplification term, which is the Holevo quantity, splits into the individual contributions from vacuum, single-photon or multi-photon signals ($\bar \chi_v$, $\bar \chi_s$ and $\bar \chi_m$).
Let us define by $p_{\mathrm{mid}}(a)$ the probability that the signal $a \in \{v, s, m \}$ sent by Alice produced a click in a middle time slot in Bob's detector, and the total probability of a middle click by $p_{\mathrm{mid}} = \sum_{a \in \{v, s, m \}} p_{\mathrm{mid}}(a)$. Then the total key rate is given by
\begin{align} \label{keyratesum}
R = \frac{1}{2} \left ( p_{\mathrm{mid}} \bar I_{\mathrm{obs}} -  \sum_{a \in \{v, s, m \}} p_{\mathrm{mid}}(a) \; \bar \chi_a \right ).  
\end{align}
A factor $1/2$ was introduced for the sifting. 

Typically, we assume that Eve has full knowledge about the vacuum and the multi-photon signals (tagging), therefore $\bar \chi_v = \bar \chi_m = 1$.  Only the term $\bar \chi_s$ enters the optimization. 
Due to the decoy and tagging methods, Alice and Bob have an estimate of the error rate within the single photon events ($q$). The error rate $q$ is the quantity appearing in the optimization and is generally different from $Q_{\mathrm{tot}}$. 

When we are no longer dealing with the strict qubit-to-qubit scenario,  we must make some modifications to the constraint on $\rho_A$ in order to adapt to the realistic scenario.  We can no longer use the full information $\rho_A$  in Eq. (\ref{rhoA}) to constrain the reduced density matrix $\tr_B (\barrh)$ in Eq. (\ref{rhoAparam}). 
Recall that the fixed reduced density matrix $\rho_A$ describes the reduced density matrix of all single photons exiting the source. These photons are lost in the channel with probability $p_{\mathrm{lost}}$, and arrive in Bob's detector with probability $1-p_{\mathrm{lost}}$. The reduced density matrix $\rho_A$ of the total photons is conserved in this process 
\begin{align}
\rho_A = (1-p_{\mathrm{lost}}) \tr_B (\barrh) +  p_{\mathrm{lost}} \rho_A^{\mathrm{lost}} \label{rhoArelaxed},
\end{align}
where $\rho_A^{\mathrm{lost}}$ is an unknown density matrix corresponding to the lost photons. 
We use this weaker version to constrain $\tr_B(\barrh)$ in the optimization of the Holevo quantity $\bar \chi_s$.


We optimize the Holevo quantity $\bar \chi_s$ over the set of symmetric states $\barrh$ (Eq. (\ref{barrhoABmat})) with the error rate constrain $q$ and the relaxed constraint in Eq. (\ref{rhoArelaxed}) on the reduced density matrix. 
The dependence of the optimized Holevo quantity on $q$,  $p_{\mathrm{lost}}$ and $\kappa$ is denoted in square brackets $\bar \chi_s^{\mathrm{max}}[q, p_{\mathrm{lost}}, \kappa]$.
The total optimized key rate for the unbalanced phase-encoded protocol is then given by
\begin{align}
R &= \frac{1}{2} \Big ( -p_{\mathrm{mid}} h(Q_{\mathrm{tot}}) + p_{\mathrm{mid}}(s)(1-\bar \chi_s^{\mathrm{max}}[q, p_{\mathrm{lost}}, \kappa]) \Big ).
\end{align}

\begin{figure} 
\includegraphics[width=\columnwidth]{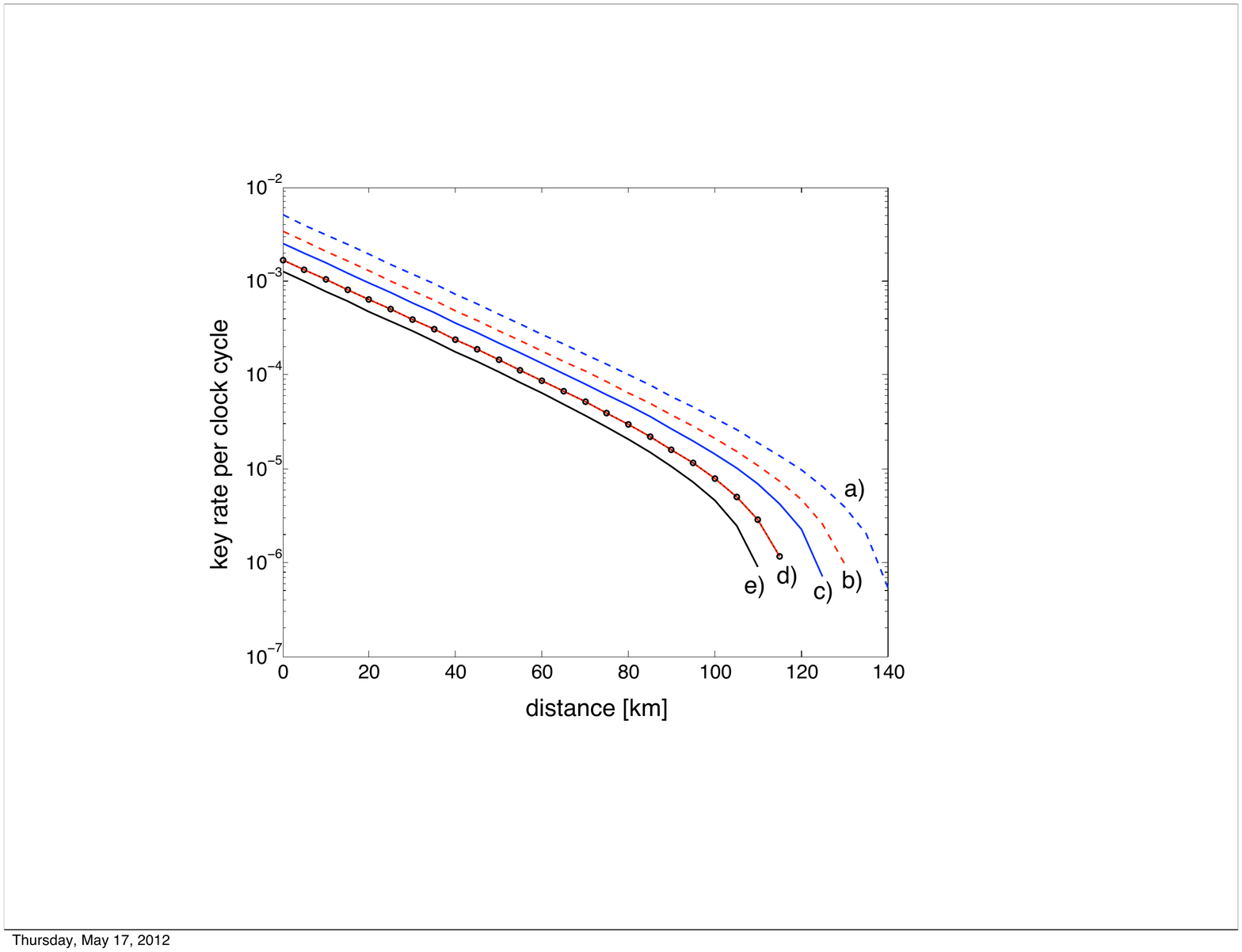}
\caption{\label{key-rate_fig} (color online) Plot of the key rates in the realistic scenario. a) Key rate of the PBS protocol with no loss (dashed blue line). b) Key rate of the PBS protocol with $\kappa=0.5$ (dashed red line). c) Key rate of the unbalanced phase-encoded protocol with no loss ($\kappa=1$) (solid blue line). d) Key rate of the unbalanced phase-encoded protocol with $\kappa = 0.5$ (solid red line) coinciding with the key rate of the hardware fix with an uneven beamsplitter (black circles). e) Key rate of the hardware fix with additional loss in the short arm (black line). }
\end{figure}

\begin{figure} 
\includegraphics[width=\columnwidth]{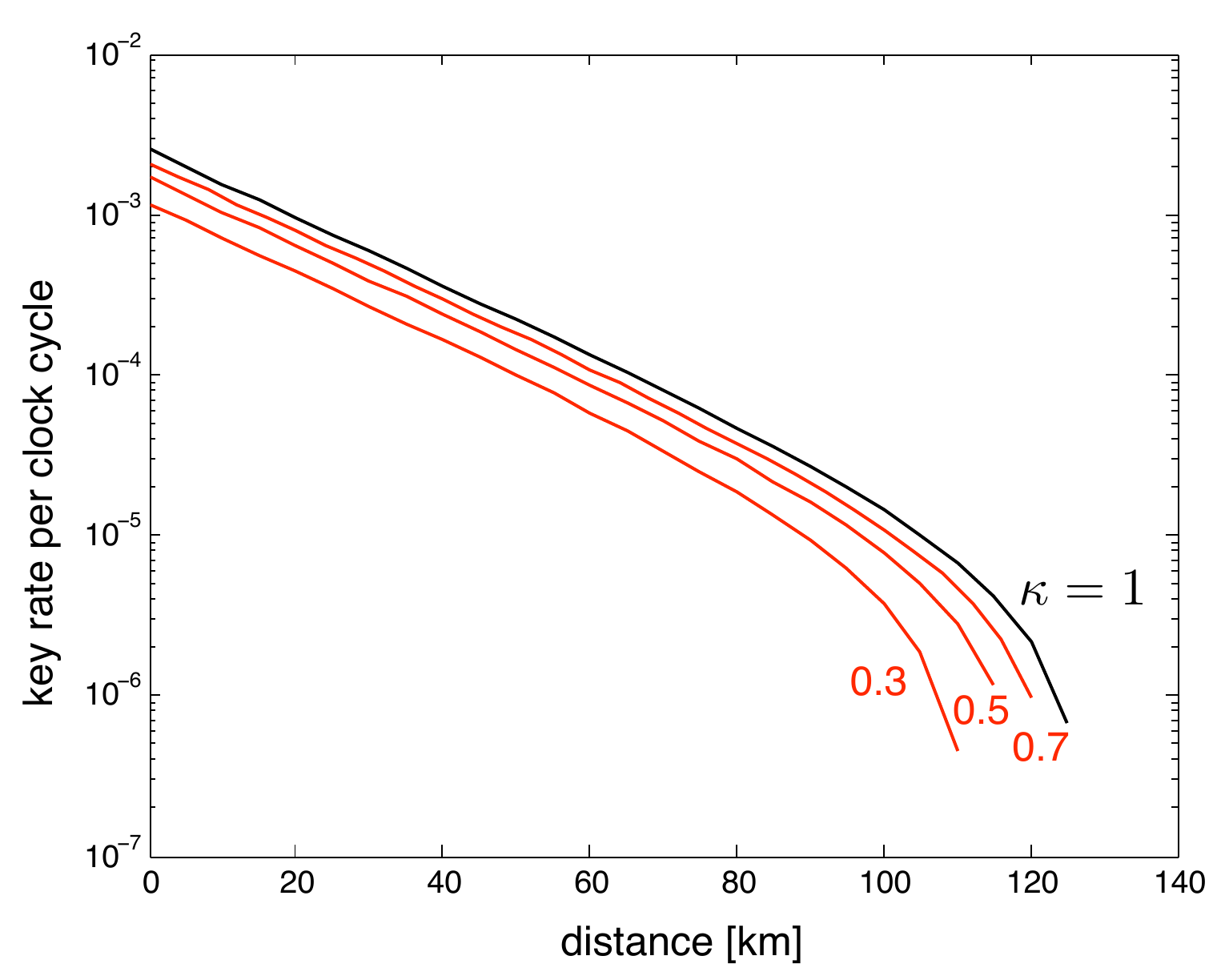}
\caption{\label{key-rate-kappa_fig} (color online) Plot of the key rates of the unbalanced phase-encoded protocol for different values of $\kappa$.}
\end{figure}

In Figs. \ref{key-rate_fig} and \ref{key-rate-kappa_fig} we plot the key rates of the unbalanced phase-encoded protocol and the hardware fixes for different values of $\kappa$. 
We simulate a channel using the experimental values in Ref. \cite{gobby04a} for channel loss, dark counts, detector efficiency and error correction efficiency, and assume that no double clicks were observed. We also optimize over the mean photon number of the signal pulses leaving Alice. 

Generally, the loss in the phase modulator decreases the key rate of the protocols. 
The performance of the unbalanced phase-encoded protocol coincides exactly with the performance of the hardware fix with an uneven beamsplitter, providing a choice between the hardware fix (requiring a special unsymmetrical beam splitter), and the improved theory solution presented here. Both of these scenarios, however, outperform the second hardware fix with an additional loss in the short arm.

\subsection{Key rate for PBS protocol with realistic devices}

For the PBS protocol detector setup the squashing map is shown to exist in Refs. \cite{beaudry08a, tsurumaru08a}. Since all pulses interfere, we can drop the specification on middle clicks in the key rate. We call the probability that Alice sent a single photon and Bob detected it by $p_{\mathrm{det}}(s)$ and the total detection probability $p_{\mathrm{det}} = \sum_{a \in \{v, s, m \}} p_{\mathrm{det}}(a)$. The key rate in the PBS scenario is then
\begin{align}
R =\frac{1}{2} \Big( -p_{\mathrm{det}} h(Q_{\mathrm{tot}}) + p_{\mathrm{det}}(s)(1- \bar \chi_s^{\mathrm{max}}[q,  p_{\mathrm{lost}} , \kappa]) \Big).
\end{align}

We plot the key rates of the PBS protocol in Fig. \ref{key-rate_fig} for different values of $\kappa$.
The key rates of the PBS protocol are higher than the key rates of the unbalanced phase-encoded key for equal loss in the phase modulator, because no signal is lost due to outside clicks. Nevertheless, the loss in the phase modulator decreases the key rates of the PBS protocol.


\section{Conclusion}

We analyze the security of the phase-encoded BB84 protocol with a lossy phase modulator in one arm of the Mach-Zehnder interferometer. We consider two protocols, the unbalanced phase-encoded and the PBS protocol. 
We provide a qubit-based security proof, which we embed in the more general framework of optical modes using the decoy states method, tagging and squashing. 

In general, it turns out that the proven secure key rates are lowered by the unbalanced loss in the Mach-Zehnder interferometer, so that a performance evaluation based on the security analysis of the standard BB84 protocol with laser pulses should not be used. The implementation with additional polarization encoding of the pulses (PBS protocol) performs better than the one with no additional polarization encoding (unbalanced phase-encoded protocol), because all signals forcedly interfere in Bob's interferometer. 
A comparison of the key rates of the unbalanced phase-encoded protocol to the key rates of the two suggested hardware fixes shows that an experimental remedy is not necessary and does not contribute to an improvement of the key rate. 

\vspace{1cm}


\section{Acknowledgements}

We thank Normand Beaudry, Xiongfeng Ma and Masato Koashi for discussion, Kevin Resch for suggesting the hardware fix with an uneven beamsplitter and Marius Oltean and David Luong for their help running the optimization program in Matlab. This work was supported by NSERC via the Discovery Programme and the Strategic Project Grant FREQUENCY, and by the Ontario Research Fund (ORF).


\bibliographystyle{unsrt}
\bibliography{qit_20120416,myown2}

\end{document}